# Introducing and Repealing an EV Subsidy: Evidence from New Zealand's Clean Car Discount

**Fabian Dunker** **Jyotsna Garg**

University of Canterbury

## Abstract

The rise in transport emissions is one of the primary concerns that has prompted the need for a transition towards cleaner transportation alternatives. Acknowledging that electric vehicles (EVs) can play a key role in curbing these transport-related emissions, New Zealand has reformed its policies over the past decade to promote their adoption. This study evaluates the impact of the financial incentives and penalties under New Zealand's Clean Car Discount (CCD), accompanied later by the introduction of the Road User Charges (RUC), on the EV adoption patterns in the country. We employ Structural Break Analysis, Regression Discontinuity Design, and a Difference-in-Differences approach to assess the impacts of these policies, using vehicle registration data from 2014 to 2024. The Structural Break Analysis and the Regression Discontinuity Design show that structural changes in vehicle registrations coincide with the timing of the policies. Using Queensland, Australia, as a control group and Newey-West standard errors, the Difference-in-Differences approach is applied to log-transformed registrations, so that the parallel trends assumption refers to equal growth rates. The introduction of the rebates coincided with a strong increase in EV registrations in New Zealand, but EV registrations grew by a similar proportion in Queensland, and the average effect relative to Queensland is not statistically significant. There is suggestive evidence that the rebates gradually accelerated EV uptake in New Zealand relative to Queensland. In contrast, the repeal of the rebates reduced EV registrations in New Zealand by about 72% relative to Queensland, although part of this decline reflects buyers bringing purchases forward before the rebates ended. Additionally, a sharp drop in average carbon emissions coinciding with the policy changes is consistent with targeted policy interventions mitigating transport-related emissions.

## 1 Introduction

Climate change is a global concern and one of the biggest challenges that nations are currently facing; see the United Nations (2025). The transportation sector is one of the greatest contributors of greenhouse gases (GHG), generating about 23% of the global energy-related CO2 emissions, where 70% of these emissions come from road transport, as reported by the IPCC (2022). With the rising emissions in the transportation sector, there has been a global trend to move towards zero-emission transport, in particular electric vehicles (EVs). As shown by Zhu et al. (2023), when coupled with renewable electricity generation, they are capable of extensively altering the transport-related emissions.

The gross emissions for New Zealand in 2023 were recorded to be 76.4 MtCO2-eq, where 41% of these emissions were carbon dioxide. Although these gross emissions have reduced by 2% (1.6 MtCO2-eq) from 2022, transport-related carbon emissions saw a rise of 5.9 MtCO2-eq since 1990, making it one of the primary sources of emissions, as reported by the Ministry for the Environment (2025b). According to NZTA (2019), the transport sector comprises around 18% of the GHG emissions in New Zealand, 90% of which comes from road transport only. Consistent with the second international climate target under the Paris Agreement, the Ministry for the Environment (2025a) notes that the New Zealand government aims to achieve 51-55% lower emissions from 2005 levels by 2035, and net zero emissions of all greenhouse gases other than biogenic methane by 2050, while facilitating a more resilient and equitable economy, as also discussed by Climate Action Tracker (2023). Encouraging clean transportation thus becomes a pivotal step in the process of meeting this ambitious proposition. As set out by the Ministry for the Environment (2022), the New Zealand government wants to replace 30% of its light vehicle fleet with Zero-Emission Vehicles (ZEVs) by 2035, such that the emissions from the on-road vehicles can be mitigated.

Over the last decade, the government has introduced several policy reforms to support the smooth transition from Internal Combustion Engine (ICE) vehicles to EVs. These were financial incentives and penalties and an expansion of the charging infrastructure powered by renewable energy. The most important policy instrument was the Clean Car Discount (CCD), which presented a dual mechanism of rebates and fees. A financial rebate was offered to incentivize consumers who bought low-emission cars, while a fee on high-emission vehicles was imposed. Additionally, light electric vehicles were exempt from the Road User Charges (RUC), a levy which is paid per km driven. Details of these policies are explained in Section 3.

The purpose of this study is to analyze and quantify the effectiveness of these policies in terms of EV penetration into the New Zealand fleet and in terms of CO2 emission. Our main data source is vehicle registration data published by the New Zealand Transport Agency (NZTA). We are particularly interested in structural breaks that occurred in this time series from 2014 to 2024. However, the presence of breaks is not sufficient evidence of a causal relationship between vehicle registrations and the policy changes, since breaks can also have other causes. To ascertain the impact of the policies surrounding financial incentives and penalties, causal inference is essential. The study uses the structural break analysis of Bai and Perron (1998) to identify any structural shifts in the EV uptake patterns and other motor technologies like Plug-in Hybrid Electric Vehicles (PHEVs), Petrol, Diesel, and Hybrid from 2014 to 2024. To quantify the immediate level shifts that occurred around each structural break, a Regression Discontinuity in Time (RDiT) approach is employed by fitting a linear regression model to each segment created by the detected breakpoints. The causal impacts of the CCD scheme on EV registrations are analyzed with a 2×2 Difference-in-Differences (DiD) regression model on log-transformed registrations. We use the Queensland, Australia, car market as a reference group for this assessment as it is a generally similar market with similar consumer preferences, and a similar regulatory framework. Because this comparison involves only two regions, inference is based on Newey-West standard errors rather than clustering, and the parallel-trends assumption is assessed with a pre-trend test. As complementary analyses, the introduction effect is also examined in a specification that allows for a change in slope, and on untransformed registrations. Since these EV initiatives were adopted to achieve lower carbon emissions by accelerating EV uptake in the country, this study also discusses the shifts in the average CO2 emissions of newly registered vehicles around the policy changes in New Zealand from 2015 to 2024.

The paper is organized as follows: Section 2 presents a literature review. Section 3 will discuss the incentive structures in New Zealand and Queensland in detail. Section 4 describes the data used for this study. The methodology used in this study is presented in Section 5. Section 6 presents the main results of the structural break analysis and the causal inference approaches. Section 7 concludes.

## 2 Literature Review

Numerous studies have been conducted to assess the structure and effects of monetary and non-monetary regulatory measures, along with the role of infrastructural development in promoting EV adoption across several countries. Sierzchula et al. (2014) in their study across 30 countries, used a multiple regression model and found that the relationship between EV adoption and the financial incentives, charging infrastructure, and local production facilities was positively correlated, whereas the sociodemographic variables were found to be insignificant. Ajanovic and Haas (2016) also identified policy incentives as important factors impacting the growth of EV uptake in these areas, while cautioning that the environmental benefits from this transition rely on the source of electricity generation. Plötz et al. (2016) employed a pooled Ordinary Least Squares (OLS) regression, which identified fiscal incentives as a significant regressor in both the United States and Europe, whereas the number of charging stations was positively correlated, thereby contributing empirical evidence to the discussion of policy-aided market evolution for EVs. Lieven (2015) suggests the combination of lower cash grants and strategic investments in charging networks is vital for increasing EV uptake. Subsequently, Rietmann and Lieven (2019) used a covariance-based structural equation model (SEM) with 2017 data from 20 nations and identified rebates, charging station density, and traffic regulations as having a positive impact on the EV market share. As per Chandra (2022), charger stock, renewable energy consumption, gasoline price, GDP per capita, mean age, and population density have a positive impact on the BEV stock but not on the PHEV. Based on a member survey of the Norwegian EV Association in June 2014, Bjerkan et al. (2016) revealed that the monetary incentives did play an important role in accelerating EV purchases in Norway. Additionally, some studies have discussed the trade-offs that come along with these concessions, favoring the EV transition. According to the study conducted by Aasness and Odeck (2015), fiscal and non-fiscal benefits for EVs increased their adoption and reduced carbon emissions in Oslo; nonetheless, the loss of toll revenues, inconvenience to public transport, and economic loss of parking fees were the challenges coupled with it. Subsequently, Halse et al. (2025), in their quasi-experimental study of the Transport Network data (2008-2017), found that local incentives proved to be a powerful instrument in increasing BEV uptake and slashing CO2 emissions, though these incentives later seemed to be favoring higher-income households. Additionally, the increasing number of on-road BEVs was the main cause of bus-lane congestion, raising traffic concerns. Gass et al. (2014) concluded that the upfront price support was effective in future cost reduction for EVs; however, this impact is dependent on their production volume in the future and highly sensitive to technical challenges. Langbroek et al. (2016) deduced fiscal incentives as a significant factor while suggesting a great alternative to direct rebates. Cansino et al. (2018) stated that rebates, free parking, exemptions from registration fees and ownership taxes, and a good charging infrastructure had a positive impact on EV adoption.

Münzel et al. (2019) observed rebates, Point of Sales (POS) tax benefits, and recurring incentives as major impact factors on the EV market in Europe between 2010 and 2017. Martins et al. (2024) also noted that the policy impact relies on GDP per capita and renewable energy resource consumption. Alali et al. (2022) also found that the financial purchase incentives were most effective when the GDP per capita change and the rebates were above average levels in the United Kingdom (UK) between 1999 and 2019. Zhang et al. (2014) used a qualitative analysis approach and identified financial incentives, technology support, and charging infrastructure as the key factors contributing to EV sales. Azarafshar and Vermeulen's (2020) study also assessed the impact of financial incentives to be greater on BEV sales than PHEV sales in Canada from 2012 to 2016. A similar outcome was later noted where purchase subsidies were deemed to be helpful in increasing EV sales and market share by Shang et al. (2024). Gong et al. (2020) deduced that discounts on energy bills and parking fees were significant measures, while direct rebates that can reduce the upfront cost of expensive EVs were the most preferred form of incentives among consumers. Lodhia et al. (2024) reiterated the importance of the presence of direct rebates and fiscal support policies to pull down the cost of EVs.

New Zealand has also adapted its EV policies as per the specific needs, scrutinized by different studies. Lemon and Miller (2013) established that while electric vehicles (EVs) can prove to be impactful devices in tackling environmental issues, at the same time they face some significant challenges to their adoption, particularly high purchase cost, short battery range, and battery life. Zhu (2016) emphasized customer preferences and market limitations while discussing the unique attributes of the electric vehicle market in New Zealand. Broadbent et al. (2021) propose that improving financial incentives, spreading fast-charging infrastructure, and raising awareness could boost demand drastically. This highlights the impact of regulatory interventions and public awareness and education initiatives on EV uptake. Hasan et al. (2021) presented an interesting comparison between the price dynamics of EVs over petrol-powered vehicles in New Zealand. They found that the Per-Kilometer Cost of Ownership (PCO) of used EVs is the lowest at 25.5 NZ cents in comparison to petrol-powered cars. Sheng et al. (2022) found that the economic development in New Zealand has a unidirectional causal relationship with the transport emissions, along with the existence of an inverted U-shape relationship between income and emissions, indicating that the emissions increase with income to a certain point before rapidly decreasing. In this study, they also established that fossil fuel consumption influences transport emissions, therefore calling for policy measures that entail decreased reliance on fossil fuels.

**Contribution to the literature** A large body of research from many countries offers evidence on the impact of fiscal benefits for EV buyers, alone or in combination with broader measures. For New Zealand, the available literature focuses on government policies, consumer choices, technological challenges, the necessary infrastructure, and the environmental impacts of EVs. However, there is a research gap regarding the causal effects of the recent policy developments in New Zealand. This study analyses both the introduction and the abrupt repeal of the CCD, which combined purchase incentives with fees on high-emission vehicles, and the subsequent introduction of RUC for EVs. The two events allow us to compare the effect of introducing a subsidy with the effect of removing it. We find that the introduction of the rebates did not increase EV registrations in New Zealand by a larger proportion than in Queensland, whereas the repeal was followed by a sharp and persistent decline, preceded by a rush of purchases before the deadline. We also discuss the changes in the average CO2 emissions of newly registered vehicles around these events.

## 3 Policy Framework: Incentives and Penalties in New Zealand and Queensland

**New Zealand** Targeted policy reforms have been enacted between 2014 and 2024 to expedite the decarbonization of road transport. The measures ranged from early discussions on the barriers of the EV market in New Zealand to cost benefits and ease of travel for EV owners, such as the exemption from RUC, a discount on ACC levy rates, and access to special lanes. Furthermore, there was financial support from the government for extending the public charging. In September 2016, further extension was given to EVs regarding the Road User Charges (RUC) exemption, on Gross Vehicle Mass (GVM) up to 3.5 tonnes, until December 2021. The Energy Innovation (Electric Vehicles and Other Matters) Amendment Act 2017 took effect, which enabled EVs access to special lanes, lower ACC Levy rates to be charged to EV owners, while an exemption to Heavy EVs from RUC was further extended too, as documented by the Ministry of Transport (2025a). The debate surrounding the Zero Carbon Bill in reaction to the Climate Change Response (Zero Carbon) Amendment Act 2019 was in support of a shift towards cleaner transport, thus initiating a modified Emission Trading Scheme (ETS) in 2020, to not only sustain but also accelerate the EV adoption, according to the Environmental Protection Authority (2025). In January 2021, another extension of the RUC exemptions for light EVs was announced until March 2024, while the Heavy EVs received the RUC exemption till December 2025, as reported by the Ministry of Transport (2025a). In July 2021, the Clean Car Discount (CCD) Scheme became a transformative step in incentivizing low-emission motor technologies while penalizing high-emission motor vehicles in New Zealand. In its first phase from July 2021, the CCD scheme provided direct rebates to buyers of new and used low-emission vehicles based on their CO2 emissions. The maximum rebates were for zero-emission vehicles (ZEVs), i.e., vehicles emitting CO2 of 0 g/km. An incentive of $8625 for a new vehicle, which was later reduced to $7015 in July 2023, and an incentive of $3450 for a used vehicle initially, which was also revised and increased slightly to an amount of $3507 from July 2023, were granted to the consumers. Rebates for these ZEVs were contingent on a price cap purchase set to $80000 (including Goods and Services Tax (GST) and on-road cost), beyond which there were no incentives offered. Similarly, new low-emission vehicles with emissions between 1-56 g CO2/km (3p-WLTP) were granted rebates of $5000, whereas used vehicles got $2000. In addition to offering rebates, the CCD scheme also imposed a fee on high-emission vehicles, starting from April 2022. Vehicles having emission levels of 192 g CO2/km (3p-WLTP) or more, were considered the high-emission level as per the Land Transport Rule: Vehicle Efficiency and Emissions Data 2022. These penalties were

capped at $4500 for new vehicles and at $2500 for used vehicles to avoid excessively high fees. The scheme also identified a third category of vehicles known to be the zero band for the vehicles emitting between 146-191 g $CO_2$/km (3p-WLTP). These vehicles neither incurred fees nor were granted any rebates. The CCD scheme, with both its rebates and fees, came to an end in December 2023, as documented by NZTA (2022). Finally, the exemption of the RUC was revoked in March 2024 for light EVs. The light BEVs and PHEVs were subject to pay RUC starting from April 2024, while heavy EVs were given an extension of RUC exemption till December 2025. The light BEVs (GVM less than 3500 kg) pay $76 per 1000 km, and the light PHEVs (GVM between 1001 and 3500 kg) pay an RUC of $38 per 1000 km, as specified by NZTA (2024).

**Queensland, Australia** Similar to New Zealand, Australia also initiated popularizing the use of EVs nationwide. The transport sector accounts for roughly 21% of Australia's emissions, and 85% of these emissions come from road transport. Therefore, with the desire to have a cleaner and greener transport system in the future, the target of 43% less emissions by 2030 and net zero by 2050 was set. Like other Australian states, Queensland also implemented EV policies as per the ZEV strategy 2022-2032 and the ZEV Action plan 2022-2024, launched in March 2022. Queensland offered AU$3000 in direct rebates for EVs purchased from 16 March 2022, when the scheme was announced, with applications opening in July 2022. This rebate applied until June 2023 for households with an annual income of up to AU$180000. These incentives were later increased in July 2023 to AU$6000, supplementing the prior recipients with an additional grant of AU$3000. In addition to EV rebates for household ownership, rebates for ZEVs up to 5 vehicles per financial year were offered to businesses with a dutiable value of up to AU$58000 (GST inclusive), excluding used vehicles, electric mobility devices, electric motorbikes, electric boats, and ZEVs above this price cap. Such direct rebates on the ZEVs ceased as of 2 September 2024. Later and to facilitate the adoption of Electric bikes and scooters, a short-term E-mobility Rebate Scheme was launched 23 September 2024 to 25 October 2024. This program offered a rebate of 500 dollars for a qualified e-bicycle and 200 dollars for a qualified e-scooter, up to 3 devices per household. Additionally, a lower stamp duty and registration fees for EVs and Hybrids were introduced, as reported by the State of Queensland (2025). Queensland thus appears to be an interesting reference point for comparison, allowing measurement of policy impact in the context of New Zealand. More importantly, New Zealand and Queensland are similar car markets with comparable consumer preferences, e.g. mid-sized pick-ups and secondhand Japanese imports have a strong market share in both regions. In addition, the standards for road legal vehicles are similar or identical in both markets. Table 1 displays the sequential comparison of the timing of the policy implementation in the two regions. The two different regions (New Zealand and Queensland, Australia) shared a very similar position regarding the EV initiatives in accelerating their EV adoption. Regarding policy implementation, though New Zealand introduced national policies, Queensland integrated federal policies with state-specific policies to aid EV uptake. However, it is to be considered that although the EV regulatory measures in Queensland were introduced in a time frame aligning with that of New Zealand, it does not exactly overlap with the timing of the treatment.

| TIMELINE | NEW ZEALAND | QUEENSLAND, AUSTRALIA |
|---|---|---|
| January 2021 | ▲ Extension Of RUC Exemptions for Light and Heavy EVs | |
| July 2021 | ▲ CCD Rebates Started | |
| March 2022 | | ▲ ZEV Rebate Scheme Launched |
| April 2022 | ▼ CCD Penalties Started | |
| July 2023 | ▼ ZEV Rebates Adjusted | ▲ ZEV Rebates Increased |
| December 2023 | ▼ End Of the CCD Scheme | |
| March 2024 | ▼ RUC Exemption Ended for Light EVs<br>▲ RUC Exemptions Continued for Heavy EVs | |
| April 2024 | ▼ RUC Implemented for Light EVs | |
| September 2024 | | ▼ ZEV Rebates Ended<br>▲ E-Mobility Scheme Launched |
| October 2024 | | ▼ The E-Mobility Rebate Scheme Ended |

**Table 1: A timeline comparison for both regions; ▲ marks the start or extension and ▼ the end or reduction of an EV incentive policy.**

## 4 Data Collection

The study presents the identification of structural breaks around policy changes, along with assessing their causal impacts. Thus, the data were collected for two sections separately.

### 4.1 Data Collection for Structural Breaks Analysis

The dataset for this part of the study was obtained from the New Zealand Ministry of Transport (MoT, Te Manatū Waka) and the New Zealand Transport Agency (NZTA, Waka Kotahi). The dataset analyzed provides comprehensive monthly records of the number of vehicles registered as per the NZTA across various motor technologies. The summary statistics for these records provide initial insights about the adoption patterns and preferences for different technologies, setting up the framework for evaluating the market trends in the presence and absence of financial incentives over the observed period from January 2014 to December 2024.

| | **Battery electric (BEV)** | **Hybrid** | **Plug-in Hybrid Electric (PHEV)** | **Petrol** | **Diesel** |
|---|---|---|---|---|---|
| Number of Observations (n) | 132 | 132 | 132 | 132 | 132 |
| Minimum | 2 | 99 | 3 | 1295 | 437 |
| First quartile | 111.2 | 381.2 | 42.25 | 10847 | 4351 |
| Mean | 650.2 | 2424.3 | 264.84 | 16175.9 | 5292.27 |
| Median | 405.5 | 1490.3 | 117 | 17638.5 | 5262.5 |
| Third quartile | 767.2 | 3744 | 310.5 | 20015 | 6152 |
| Maximum | 4524 | 11037 | 1611 | 30302 | 13129 |
| Range | 4522 | 10938 | 1608 | 29007 | 12692 |
| Standard deviation (sd) | 823.54 | 2534.49 | 350.3 | 5249.39 | 1468.4 |
| Skewness | 2.19 | 1.19 | 1.9 | -0.51 | 0.75 |
| Kurtosis | 4.89 | 0.73 | 3.11 | -0.47 | 5.58 |
| Standard error (se) | 71.68 | 220.6 | 30.48 | 456.90 | 127.81 |

**Table 2: Summary statistics for different motor technologies from January 2014 to December 2024.**

### 4.2 Data Collection for Difference in Differences

The comparison dataset for the difference in differences analysis was obtained from the Department of Transport and Main Roads (TMR), Queensland, Australia, for different fuel types. To conduct this analysis, a few data transformations were done to both New Zealand's and Queensland's vehicle registration data to achieve credible results. The variable of interest for this study is the EV registrations in both regions. The New Zealand data are initial registrations, whereas the Queensland data are the number of registered vehicles at the end of each month, which is close to the fleet size. The Queensland series was therefore differenced once as per the following equation.

$$\Delta\ Registrations = (Registrations)_t - (Registrations)_{t-1} \quad , \text{where } t \text{ represents time}$$

Negative values, which occur when more vehicles leave the register than are added, were replaced by 0. For Queensland, the EV series is the fuel type "Electric" summed over all vehicle types except the category "Other"; it therefore includes electric motorcycles but no plug-in hybrids. For New Zealand, the registrations of BEVs and PHEVs were merged into a single variable. The two EV series are therefore not defined in exactly the same way, which should be kept in mind when interpreting the results.

## 5 Methodology

### 5.1 Structural Break Analysis

The Bai and Perron (1998) methodology for structural breaks considers a linear regression model that assumes that the relationship between the regressors and the dependent variable is stable within the segment but tends to change across the unknown breakpoints. The break dates are estimated by minimizing the sum of squared residuals over all possible partitions of the series, and the number of breaks is selected with the Bayesian Information Criterion (BIC). The break date estimators are consistent under general conditions, and Bai and Perron (1998) also allow for heteroscedasticity and autocorrelation in the errors. In this study, the model within each segment is a constant mean, so that the breaks correspond to shifts in the level of the series. The focus is on the location of the breaks; the analysis is descriptive and does not by itself identify causal effects.

The analysis of structural breaks in the number of registrations is conducted using the strucchange package in R (Zeileis et al., 2002). The minimum segment length h was set to 10 months, to ensure a sufficient number of observations for a reliable estimation.

### 5.2 Regression Discontinuity in Time Analysis

After the identification of the breakpoints, we quantify the size of the changes at each break. We use the regression discontinuity in time (RDiT) framework introduced by Hausman and Rapson (2018), to measure the level shifts around the introduction and removal of the financial incentives. Here, we intend to assess the shifts at the identified breakpoints so as to measure the immediate changes around the policy interventions by fitting a linear regression model separately to each detected segment. A separate linear regression model was fitted to the segments before and after each structural change. Trends and baseline levels within each segment were visualized using regression lines. This was measured by calculating level shifts at each breakpoint, defined as the difference between the predicted values immediately before and after the breakpoints.

$$\text{Level pre break} = \hat{\beta}_i + \hat{\gamma}_i \times \text{breakpoint time},$$

$$\text{Level post break} = \widehat{\beta_{i+1}} + \widehat{\gamma_{i+1}} \times \text{ breakpoint time.}$$

The level shift was computed as:

$$\text{Level Shift} = \text{Level post break} - \text{Level pre break}$$

Where $\widehat{\beta_i}$ and $\widehat{\beta_{i+1}}$ are the estimated intercepts for segments $i$ and $i + 1$, and $\widehat{\gamma_i}$ and $\widehat{\gamma_{i+1}}$ are the estimated slopes for segments $i$ and $i + 1$, while the *breakpoint time* is the time of the structural break. This quantifies the immediate change at each break. Since the breaks are estimated from the same data and other events may coincide with them, these level shifts are descriptive rather than causal.

### 5.3 Carbon Emissions Forecast

The structural breaks in the monthly average carbon emissions were identified with the strucchange package (Zeileis et al., 2002), and the trend in each segment was extrapolated until December 2027. The predictions were done using the forecast package in R by Hyndman and Khandakar (2008). To investigate further, the annual carbon emissions per vehicle were calculated

$$\text{carbon emissions} = (\text{average carbon emissions} \times \text{annual distance traveled})/1000000.$$

The carbon emissions on the left-hand side are the annual emissions per vehicle (in tonnes). On the right-hand side, the average carbon emissions are the monthly average emissions of newly registered vehicles (in g/km), either actual or extrapolated, and the annual distance traveled is the average distance per vehicle in New Zealand. This gives the annual emissions that a vehicle registered in that month would contribute. The extrapolations are mechanical continuations of the segment trends and are only meant to illustrate the differences between the segments.

### 5.4 Difference-in-Differences Approach

This approach compares New Zealand, the treated group, and the control group, i.e., the group that did not receive the treatment, in the context of the evolving EV adoption before and after the introduction of the rebates. This quasi-experimental design makes use of the monthly data of both regions and mitigates the biases emerging from time-invariant unobserved heterogeneity and other confounding factors influencing the EV adoption in New Zealand post-policy intervention. Difference-in-Differences (DiD) is a widely employed statistical method for estimating the impact of a policy intervention on a group exposed to that policy change. It compares the changes in the outcome variable, or dependent variable, of the treated group, with the changes in the outcome variable of the group devoid of any treatment, called the control group. By differencing the average change in the dependent variable of the control group and the treated group before and after the treatment, it estimates the average treatment effect of the policy, which can be interpreted as a causal effect if the assumption described next holds. This method relies on the presumption that, in the absence of treatment, the outcome for the two groups would have followed similar trajectories. This assumption, known as the parallel trends assumption, is the identifying condition for the DiD estimator. It is important to emphasize that parallel trends is a statement about an unobserved counterfactual and is therefore not directly testable. Its credibility can, however, be assessed by examining whether the treated and control series moved in parallel during the pre-treatment period, and the robustness of the estimates can be examined with alternative specifications; both are undertaken below.

To quantify the impact of these EV incentives that New Zealand was exposed to on its EV adoption, as compared to Queensland, this analysis will aid in the evaluation of the average change in EV registrations in New Zealand before and after the introduction of subsidies, when differenced with the average change in Queensland's EV registrations, exposed to no policy intervention in the time frame considered. For this study, the outcome variable is the logarithm of the monthly EV registrations for New Zealand and Queensland, and the treatment is considered to be the EV incentives offered in New Zealand under the CCD scheme from July 2021 to December 2023. A regression-based 2×2 DiD framework was implemented, with time windows chosen around the policy changes. Table 3 represents the 2×2 DiD table:

| | **Pre-treatment** | **Post-treatment** | **First difference** |
|---|---|---|---|
| Control group | $Y_{11}$ | $Y_{12}$ | $Y_{12} - Y_{11}$ |
| Treated group | $Y_{21}$ | $Y_{22}$ | $Y_{22} - Y_{21}$ |
| DiD estimator | | | $(Y_{22} - Y_{21}) – (Y_{12} - Y_{11})$ |

**Table 3: The 2×2 Difference-in-Difference estimate calculations.**

For a feasible application of DiD in this case, two separate timelines were considered, omitting the period from March 2022 to July 2023, during which Queensland introduced its own rebate and New Zealand extended and revised the CCD scheme. The rebates in New Zealand started in July 2021, whereas purchases in Queensland became eligible for its rebate from 16 March 2022. The first model therefore compares a pre-treatment period from July 2019 to June 2021 with a post-treatment period from July 2021 to February 2022. The pre-treatment period starts in July 2019 because the Queensland series contains months with very few registrations before that date, which makes the log-transformed series unreliable, and because the log gap between the two regions is approximately stable from that date onwards. The second model compares August to December 2023 with January to December 2024, to capture the change in EV adoption after the discontinuation of the EV incentives in New Zealand in December 2023, as compared to Queensland. In both models, the outcome is the logarithm of monthly EV registrations. Because the New Zealand market was about four times larger than the Queensland market before 2021, equal growth rates are a more plausible counterfactual than equal absolute changes, so the parallel trends assumption is stated in terms of growth rates. The DiD regression model is given by

$$y_{it} = \beta_0 + \beta_1 \times \text{treated}_i + \beta_2 \times \text{post}_t + \beta_3 \times (\text{treated}_i \times \text{post}_t) + \varepsilon_{it}.$$

Here $y_{it}$ is the logarithm of the number of EV registrations for region i (New Zealand or Queensland) in month t. The variable $\text{treated}_i$ is 1 for New Zealand (treated group) and 0 for Queensland (control group). In the introduction model, $\text{post}_t$ is 1 from July 2021 to February 2022 and 0 from July 2019 to June 2021; in the repeal model, $\text{post}_t$ is 1 from January to December 2024 and 0 from August to December 2023. The interaction term $\text{treated}_i \times \text{post}_t$ is the DiD term. In this equation, $\beta_0$ is the average log number of monthly EV registrations in Queensland in the pre-treatment period, $\beta_1$ is the average difference in log registrations between New Zealand and Queensland in the pre-treatment period, and $\beta_2$ is the change in Queensland's average log registrations between the two periods. Finally, $\beta_3$ is the DiD estimator, which measures how much more the average log registrations changed in New Zealand than in Queensland, so that $\exp(\beta_3) - 1$ approximates the relative effect of the policy, while $\varepsilon_{it}$ is the error term. As the model contains no time trends, all coefficients refer to changes in the average level of the log series rather than in slopes.

Two features of the present design warrant particular care in inference. First, with only two units cluster-robust standard errors by region are not appropriate, since valid clustering requires a large number of clusters. The principal threat to inference here is instead the serial correlation inherent in monthly registration series (Bertrand et al. 2004). Accordingly, all standard errors reported for the DiD estimates are heteroscedasticity- and autocorrelation-consistent (HAC) Newey-West standard errors, computed on the monthly series, with the lag length set by the common rule of thumb as the integer part of $4(n/100)^{2/9}$, where n is the number of monthly observations in the regression. Second, because the parallel trends assumption underpins the causal interpretation of the DiD estimator, its plausibility is assessed directly for each event. A pre-trend test is conducted by regressing the difference between the log New Zealand and log Queensland monthly registration series (the log gap) on a linear time trend over the corresponding pre-treatment window: a slope statistically indistinguishable from zero is consistent with parallel trends, whereas a non-zero slope indicates a pre-existing divergence that the DiD estimator would erroneously attribute to the policy. As complementary analyses for the introduction event, a specification that allows for a change in the level and in the slope of the log gap at the start of the policy is estimated, and the analysis is repeated on untransformed registrations, where a group-specific linear trend is included (Angrist and Pischke, 2009). Because a trend estimated over the full sample can absorb part of the policy effect (Wolfers, 2006), such trends are estimated from the pre-treatment period only.

## 6 RESULTS AND DISCUSSION

### 6.1 Structural Shifts in Electric Vehicle Adoption Patterns

This section presents the identification and visualization of the structural breaks according to the Bai and Perron (1998) methodology. The monthly registration data for all motor technologies (BEV, PHEV, Hybrid) have high variation; looking solely at the original time series might be misleading. Hence, for identifying the location of structural breaks in these uptake patterns, it was crucial to stabilize these variations and reduce the impact of extreme values by applying the log-transformation.

Figure 1 depicts the identified structural breaks along with the 99% confidence intervals in the BEVs, PHEVs, and Hybrid motor technologies' registration trends from 2014 to 2024, capturing the breakpoints in the uptake of these low-emission technologies around pre-, post-, and policy intervention times.

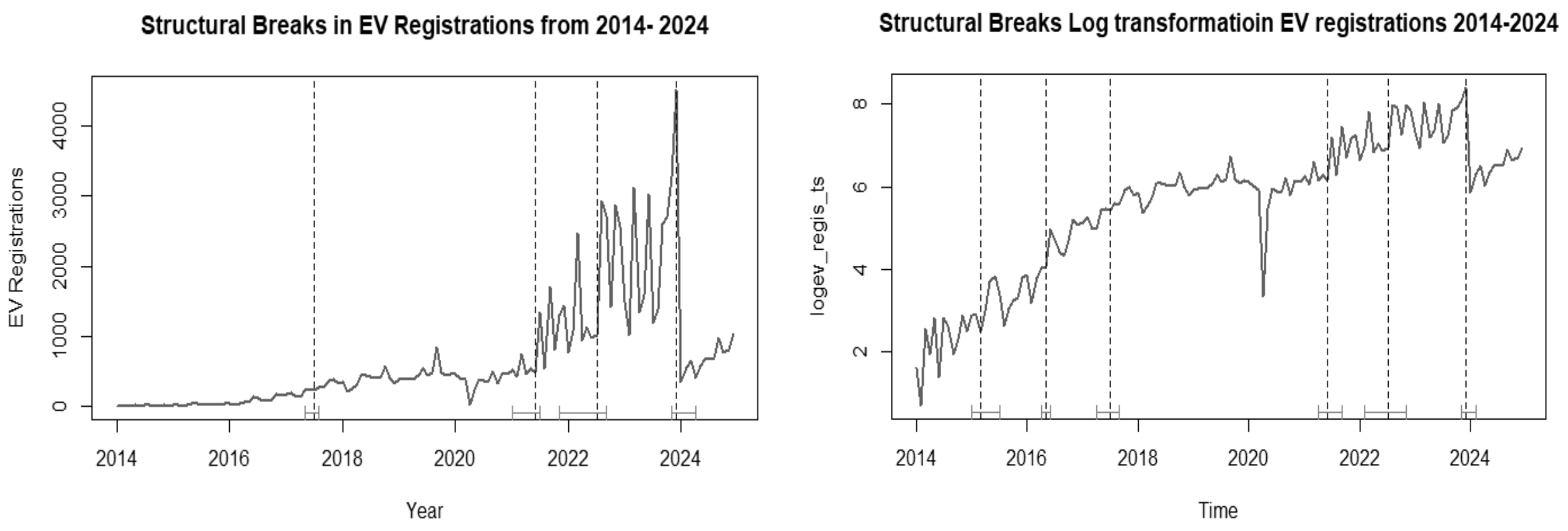

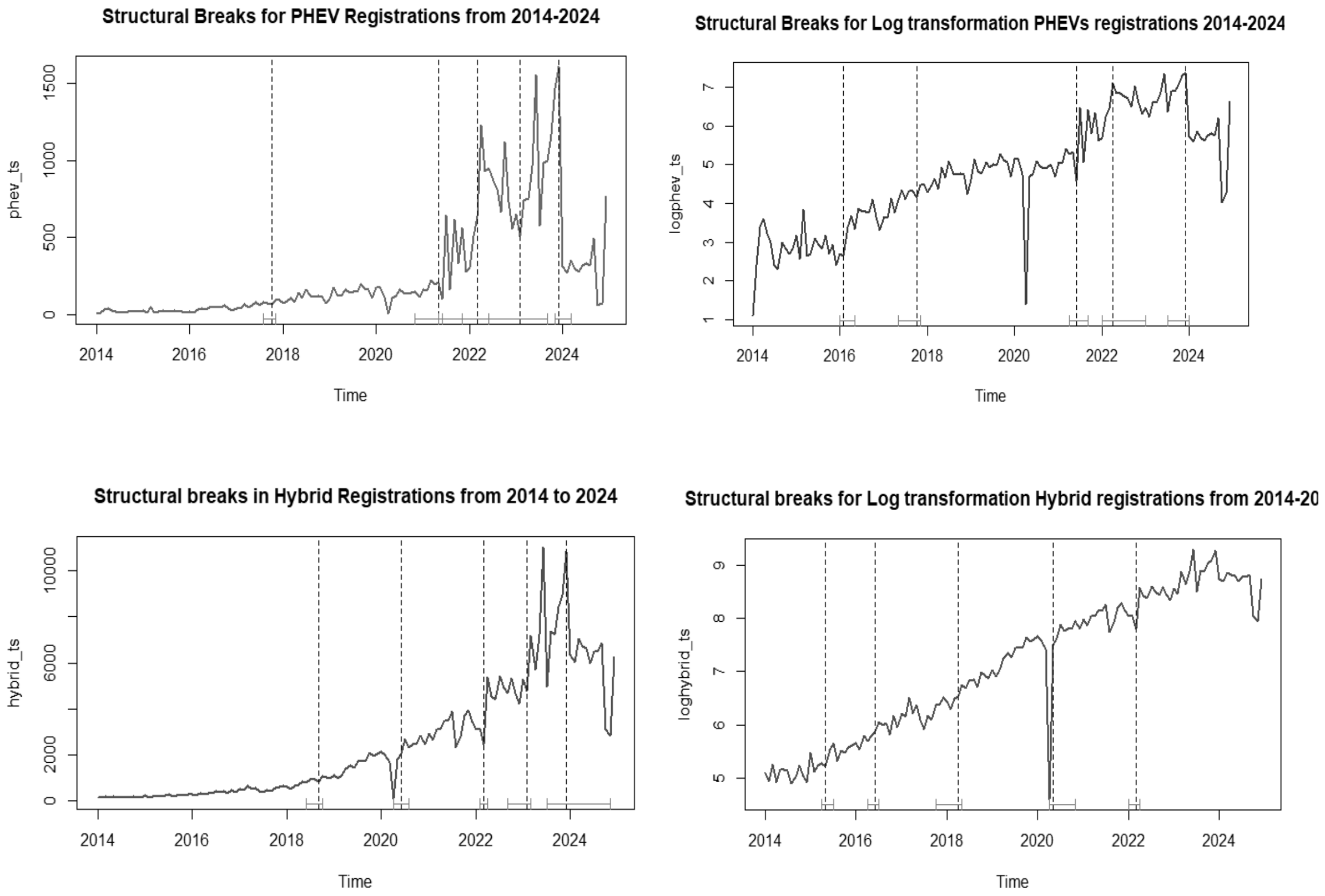

**Figure 1: Structural breaks in Original and log-transformed vehicle registrations (EV, PHEV, and Hybrid) from 2014 to 2024.**

| Motor Technologies | Break dates for the original time series | Break dates for log-transformed time series |
|---|---|---|
| Battery Electric Vehicles (BEVs) | July 2017<br>June 2021<br>July 2022<br>December 2023 | March 2015<br>May 2016<br>July 2017<br>June 2021<br>July 2022<br>December 2023 |
| Plug-in Hybrid Vehicles (PHEVs) | October 2017<br>May 2021<br>March 2022<br>February 2023<br>December 2023 | February 2016<br>October 2017<br>June 2021<br>April 2022<br>December 2023 |
| Hybrid Vehicles | November 2018<br>June 2020<br>March 2022<br>February 2023<br>December 2023 | May 2015<br>June 2016<br>April 2018<br>May 2020<br>March 2022 |

**Table 4: Break dates for original and log-transformed time series (2014–2024)**

In Figure 1, the most prominent breaks in the original time series for vehicle registrations are identified between the years 2021 and 2023. Therefore, the sharp dip in the registrations of BEVs, PHEVs, and Hybrids in December 2023 emerges as the consistent breakpoint in all three time series. There are some additional breakpoints identified in the log-transformed time series, possibly due to external shocks or market transformations driven by improved vehicle technology or better infrastructure for these motor technologies. The breaks identified for the BEVs and the PHEVs align closely with the policy timeline, which suggests that they are related to the policy changes around 2021 to 2024. The introduction of financial incentives in July 2021 and the withdrawal of the rebate scheme in December 2023 coincided with the acceleration and deceleration of the adoption of cleaner vehicles, where the number of monthly registrations for BEVs and PHEVs gained momentum during this time but saw a sharp decline after the withdrawal of the CCD. Simultaneously, Hybrids seem to be showing steady growth throughout the period, though their adoption did have a structural change around March 2022, with a peak number of registrations shortly before the CCD fees and the extended rebates started in April 2022. Their adoption does not appear to be impacted as much as their alternatives in the absence of rebates, thus maintaining their share of the fleet.

There is also a discernible but brief dip in 2020, which can be observed in all series. This dip can be attributed to the lockdown restrictions due to the COVID-19 pandemic, where there were almost no sales or registrations for a few months. However, to identify the structural changes in the adoption, a 10-month window between two consecutive structural breaks was considered here. Therefore, the lockdown period did not seem to have a sufficient number of observations to provide reliable evidence of the presence of a structural break. This sudden decline was most likely short enough not to be identified as a structural break in the BEV and PHEV series, as these series rebounded to their earlier patterns just after the dip. For Hybrid, Petrol and Diesel vehicles, however, breaks were identified in 2020 (Tables 4 and 6), which most likely reflect the lockdown. It is also interesting to observe in Figure 1 that, unlike the original time series for the Hybrid Vehicles, there is no structural break identified at December 2023 in the log-transformed series. This possibly indicates that the dip for Hybrids after the discontinuation of CCD was not as significant as the BEVs and the PHEVs.

### Structural Breaks in New and Used BEVs' Adoption Patterns

The CCD distinguished between the rebates paid for new and used BEVs. Since the New Zealand automobile market has a significant demand for secondhand or imported vehicles, it becomes essential to analyze the structural changes in the registration for both new and used BEVs.

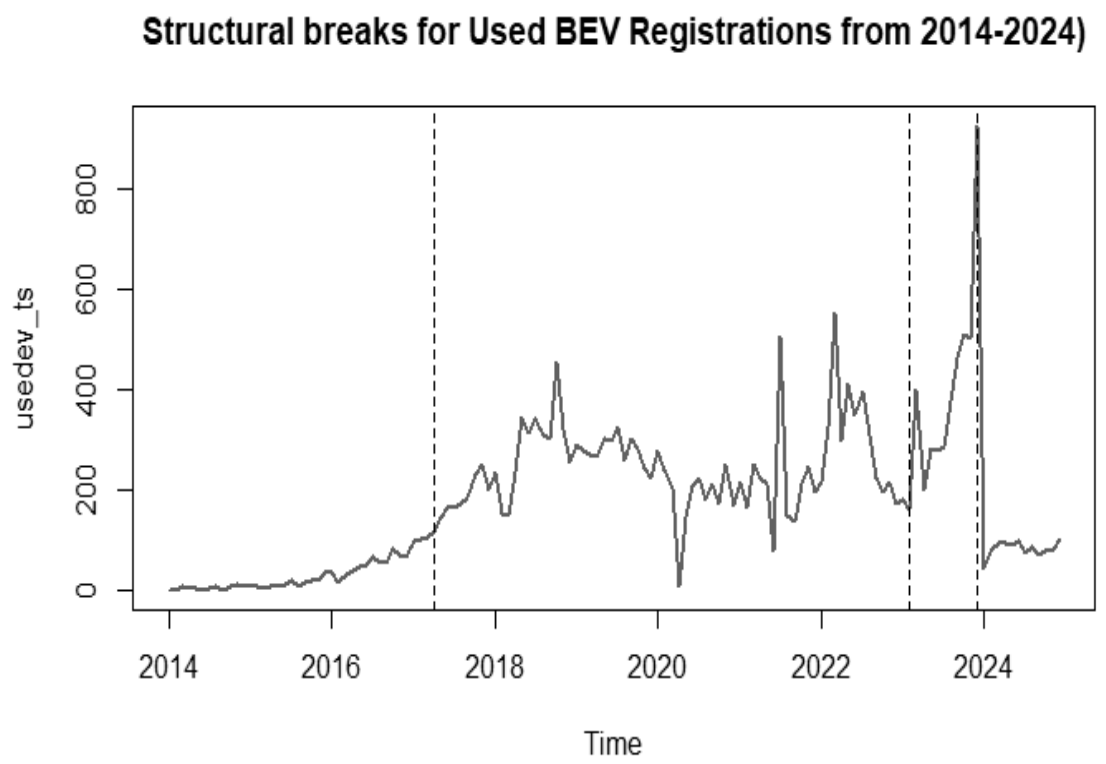

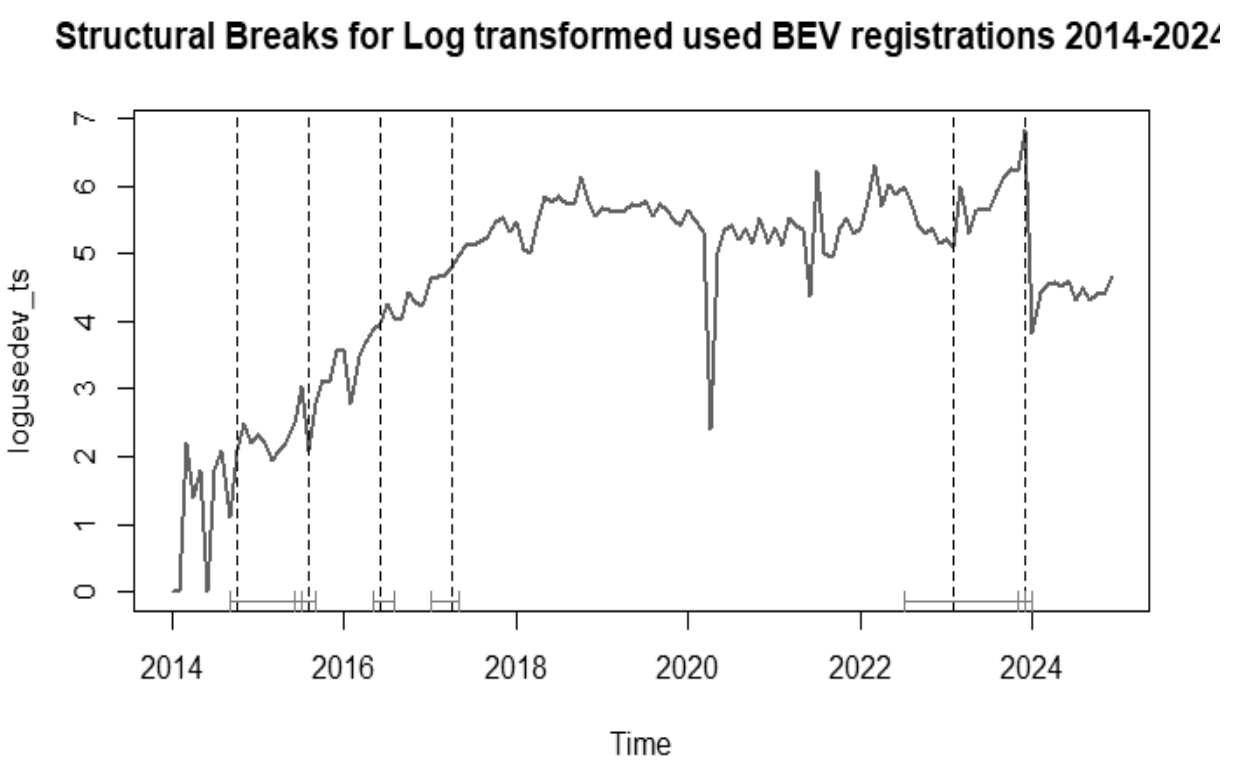

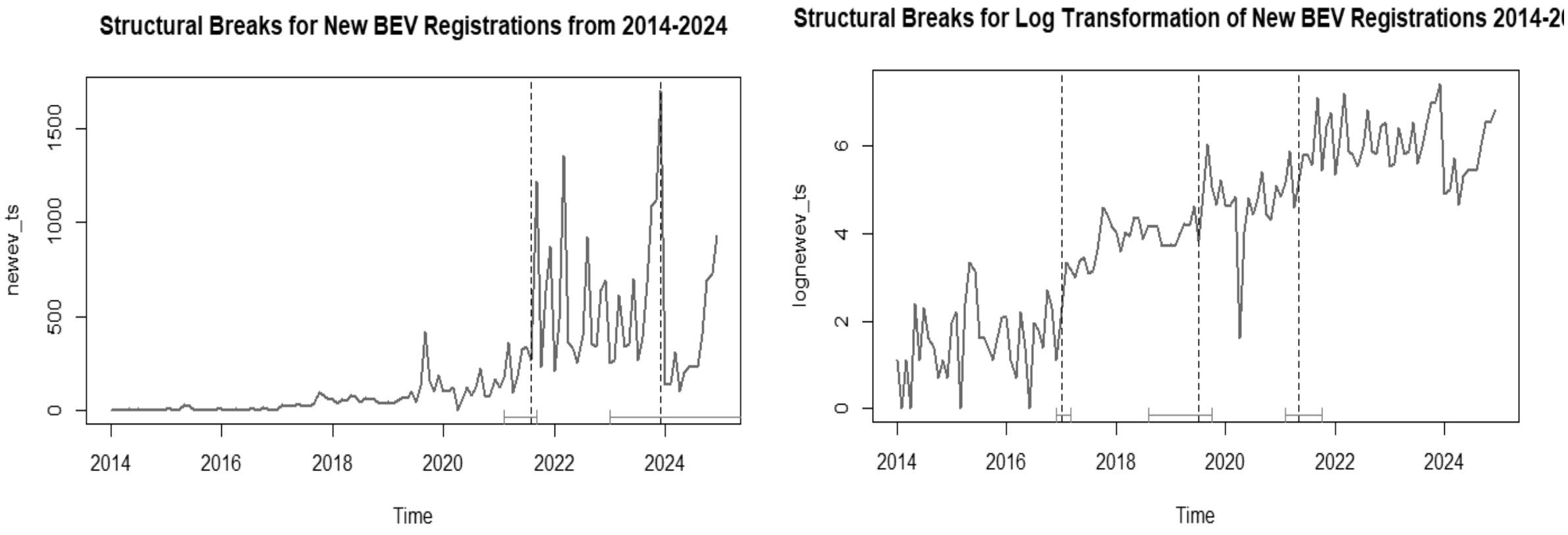

**Figure 2: Structural breaks in Original and log-transformed vehicle registrations for New and Used BEVS from 2014 to 2024.**

| Battery Electric Vehicles (BEVs) | Break dates for Original Time Series (2014–2024) | Break dates for Log-Transformed Time Series (2014–2024) |
|---|---|---|
| Used BEVs | April 2017<br>February 2023<br>December 2023 | October 2014<br>August 2015<br>June 2016<br>April 2017<br>February 2023<br>December 2023 |
| New BEVs | August 2021<br>December 2023 | January 2017<br>July 2019<br>May 2021 |

**Table 5: Break dates for Battery Electric Vehicles (BEVs): Original vs. Log-Transformed Time Series (2014–2024)**

In Figure 2, the monthly registration data for the new BEVs, apparently follows the same trends as the EV adoption in Figure 1 and thus identifies similar structural breaks in their original time series, aligning with the removal dates of the CCD. In contrast, its log-transformed time series seems to detect some different structural breaks. It is observed that the new BEVs have consistently made organic growth all these years, although their growth in the market did stabilize after May 2021. This persistent adoption pattern was probably supported by the significant rebates that pulled their prices down during that period. Furthermore, the used BEVs identified similar structural breaks as the EVs in Figure 1, where it did experience a sharp decline in registrations after December 2023, which may indicate a greater impact of the government interventions on their uptake as compared to the new BEVs. Whereas their log-transformed series identified some additional structural breaks prior to 2017, this could possibly be attributed to the availability of cheap secondhand EV models in the market. It reflects the growing popularity of used BEVs prior to 2017, evident through the constant shifts in the adoption trend that stabilized later.

### 6.2 Structural Shifts in Traditional Fuel Vehicles

While the EV uptake trends did see some visible structural changes, the conventional fuel vehicle presented an interesting parallel case as well.

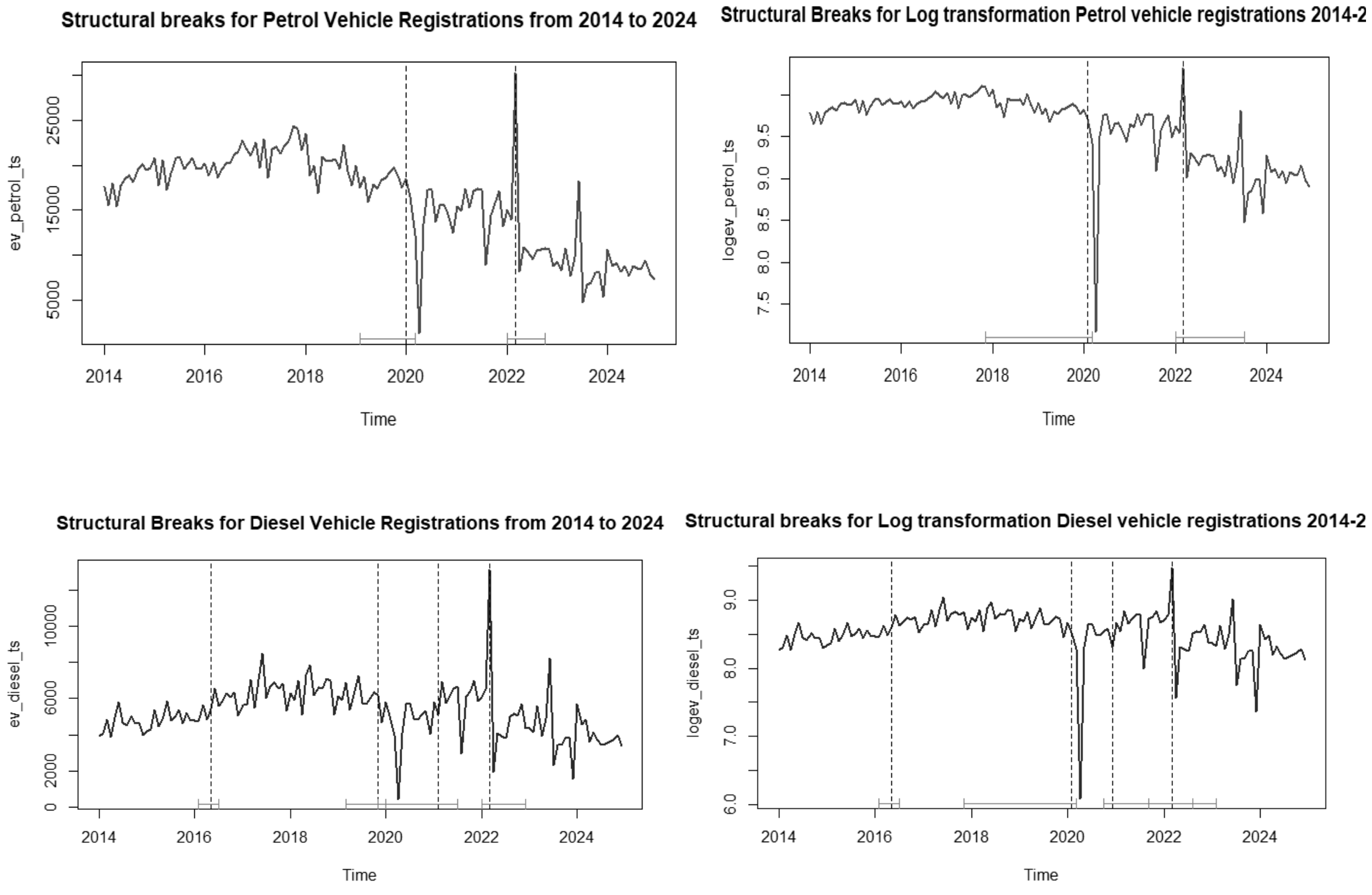

**Figure 3: Structural breaks in the original and log-transformed vehicle registrations for Petrol and Diesel vehicles from 2014 to 2024.**

| Motor Technologies | Breakdates For Original Time Series (2014-2024) | Breakdates For Log Transformed Time Series (2014-2024) |
|---|---|---|
| Petrol Vehicles | January 2020<br>March 2022 | February 2020<br>March 2022 |
| Diesel Vehicles | May 2016<br>November 2019<br>February 2021<br>March 2022 | May 2016<br>February 2020<br>December 2020<br>March 2022 |

**Table 6: Breakdates for Petrol and Diesel Vehicles: Original vs. Log-Transformed Time Series (2014–2024)**

In contrast to the BEVs and PHEVs, fewer breakpoints in the petrol and diesel vehicle registration were identified. The March 2022 breakpoint appears to be the omnipresent structural break in both the original and the log-transformed time series, exactly preceding the start of the CCD fees. This suggests a strong temporal alignment between the presence of a structural change and the policy intervention in April 2022, where people might have rushed to purchase vehicles with high emissions prior to the imposition of penalties. Although it seemed like it experienced some more peaks and troughs after the breakpoint, it still maintained a steady position in the market. Unlike the implementation of CCD, the withdrawal of the scheme in December 2023 did not cause any structural changes in the adoption patterns of both Petrol and Diesel vehicles, and they seem to continue to maintain a large share in the fleet every month. These patterns also suggest that, though the registrations for these vehicles had a significant drop after the implementation of the penalties, the removal of these penalties did not suddenly boost their registrations. Therefore, unlike the high variations in the EVs' adoption patterns, the traditional fuel vehicles observed relatively stable uptake over time.

## Structural shifts in the registrations of Utes

Ute (utility vehicles) is the Australian and New Zealand name for pick-up trucks. In particular mid-sized pick-ups represent a large market segment with a market share of around 20%. These high-emission vehicles are popular with both businesses and individuals. Figure 4 depicts the monthly registrations of these Utes and the structural breaks identified in their adoption patterns for both the original time series and the log-transformed series. This highlights shifts in their uptake around the implementation and withdrawal of the penalties.

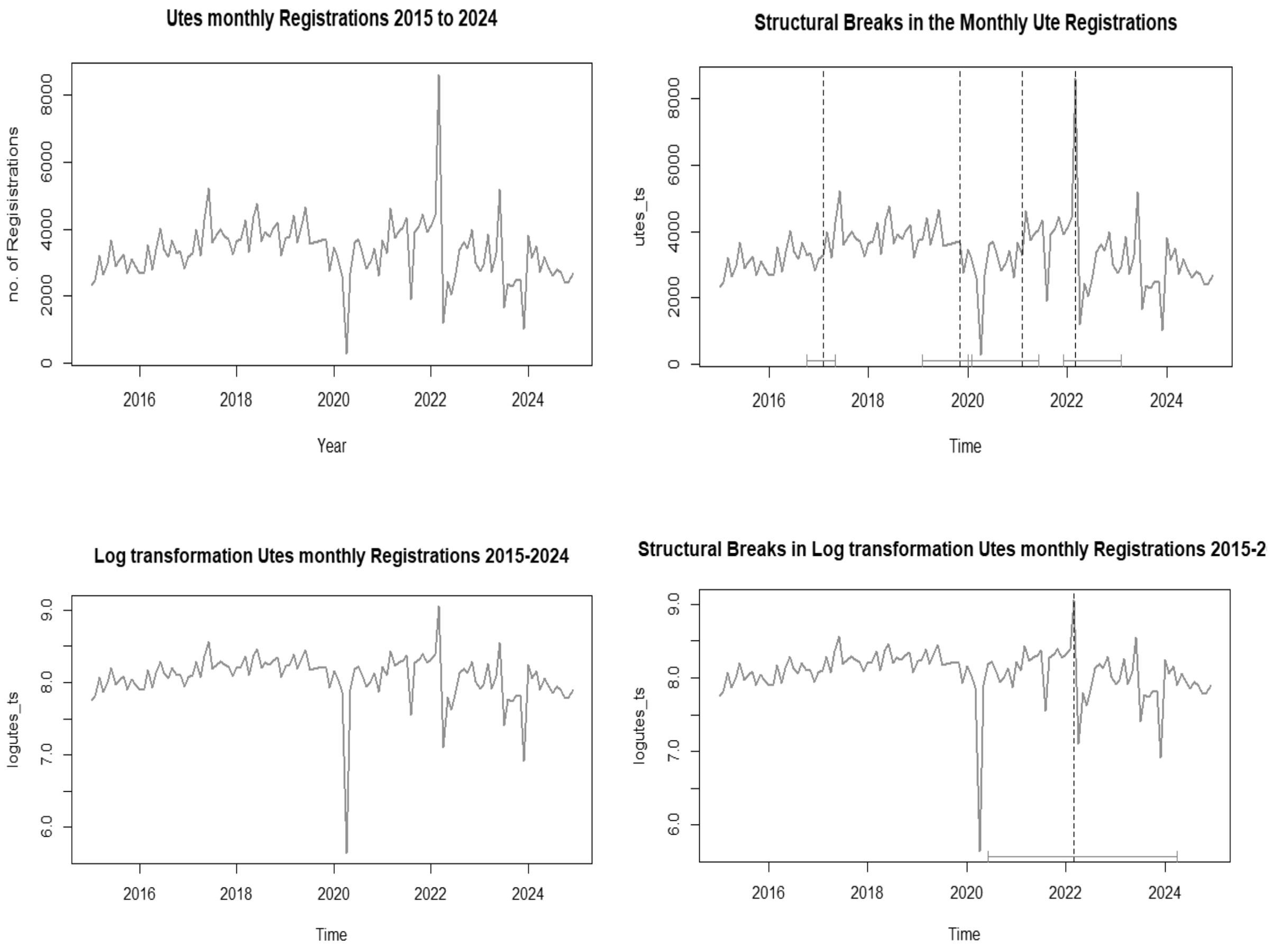

**Figure 4: Structural breaks in Original and log-transformed Utility vehicle registrations from 2015 to 2024.**

| Motor Technologies | Break Dates for Original Time Series (2015-2024) | Break Dates for Log-Transformed Time Series (2015-2024) |
|---|---|---|
| Utes | February 2017<br>November 2019<br>February 2021<br>March 2022 | March 2022 |

**Table 7: Break Dates for Utes: Original vs. Log-Transformed Time Series (2015–2024)**

Figure 4 presents a similarity in the adoption patterns of Utes compared to Petrol and Diesel vehicles. The monthly registrations of Utes not only seem to imitate the patterns but also identify resonating structural breaks. The structural break of March 2022, immediately before the penalty imposition, reiterates that these policy shifts did boost the Ute adoption for a short period, followed by a transitory decline that stabilized later. In contrast to the high variations in the EV uptake, owing to their popularity, the Utes sustained their relevance in the fleet throughout. Apart from a noticeable structural change in their patterns in March 2022, Utes, despite the policy interventions, did not observe any other drastic drop. Nevertheless, these high-emission vehicles faced increased competition in later years from alternatives whose prices were lowered by the rebates. These changes in the patterns also appear to support the notion that the revocation of the rebates had a far greater impact on the monthly registration of different motor technologies as compared to the withdrawal of penalties. This is consistent with the policy changes being associated with visible structural changes in EV adoption.

### 6.3 Magnitude of Structural Breaks Identified

In this section, we quantify the level and slope shifts at the breakpoints.

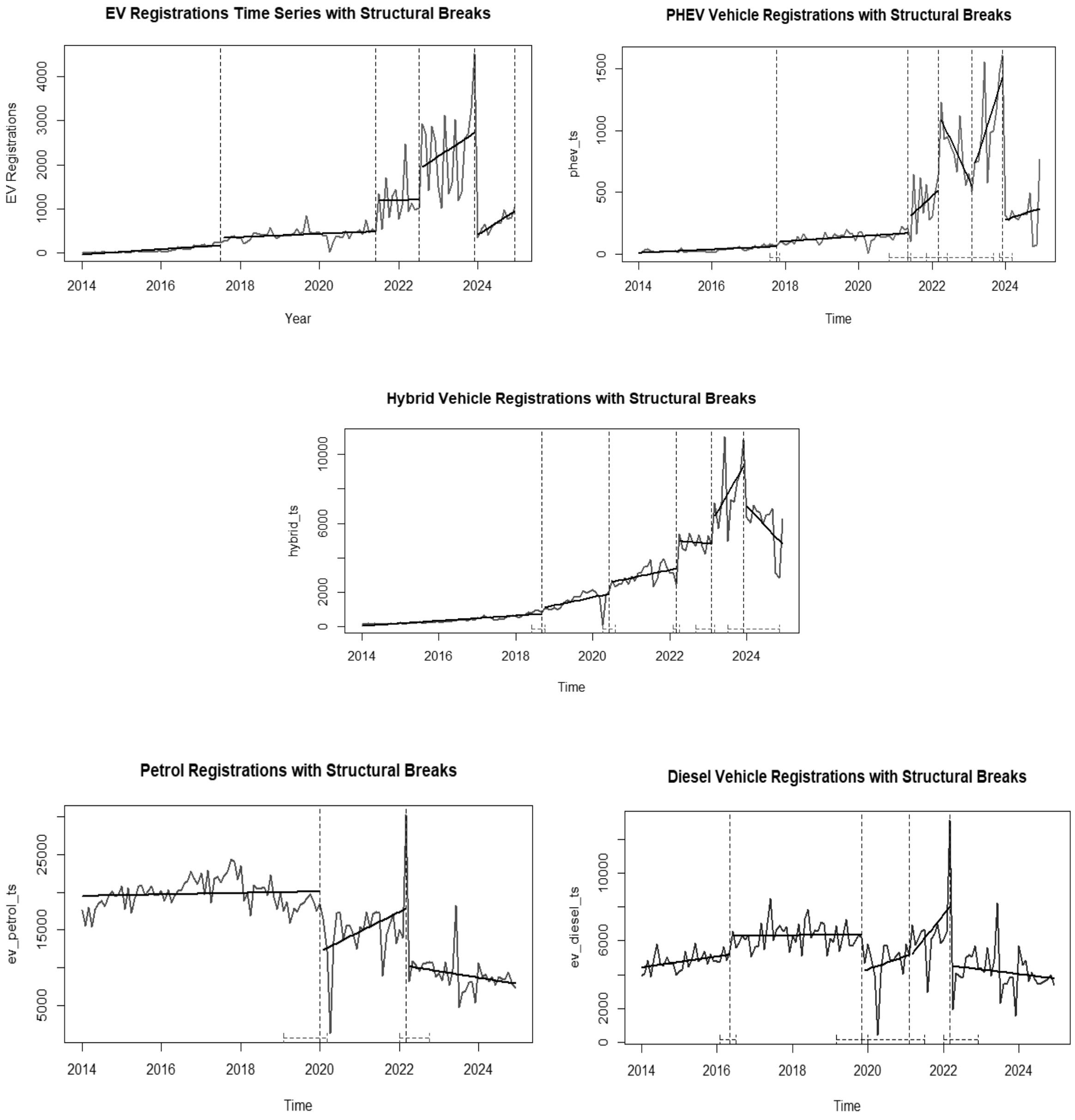

**Figure 5: Regression discontinuity models for monthly vehicle registrations (2014–2024) for Motor Technologies.**

| Motor Technologies | Break dates | Level Shifts (vehicles registered per month) |
|---|---|---|
| Battery Electric Vehicles (BEVs) | July 2017 | 165.76 |
| | June 2021 | 694.19 |
| | July 2022 | 697.55 |
| | December 2023 | -2356.85 |
| Plug-in Hybrid Vehicles (PHEVs) | October 2017<br>May 2021<br>March 2022<br>February 2023<br>December 2023 | 37.2<br>121.03<br>622.07<br>107.42<br>-1154.3 |
| Hybrid Vehicles | November 2018<br>June 2020<br>March 2022<br>February 2023<br>December 2023 | 345.89<br>653.69<br>1594.85<br>1338.97<br>-2138.58 |
| Petrol Vehicles | January 2020<br>March 2022 | -7914.98<br>-7637.79 |
| Diesel Vehicles | May 2016<br>November 2019<br>February 2021<br>March 2022 | 1084.49<br>-2175.36<br>-178.13<br>-3500.5 |

**Table 8: Break Dates and Associated Level Shifts in Monthly Registrations (2015–2024)**

Table 8 consists of level shifts in the monthly registrations corresponding to their identified breakpoints, while Figure 5 provides a visual representation of these shifts, rendering a more discernible observation of the regression discontinuities between different segments. From the plots for BEVs, PHEVs, and hybrid vehicles, significant jumps and drops are observed. The largest positive level shifts occur at the 2021 and 2022 breakpoints, around the start of the CCD and its extension in April 2022. Whereas the large negative level shift after the December 2023 breakpoint represents the combined effect of CCD withdrawal and RUC announcements. Hence, it suggests that there was previously a marked surge for EVs that suddenly stopped after the cancellation of rebates and applicability of road user charges. Simultaneously, it is noteworthy that vehicles operating on conventional fuel engines, i.e., petrol and diesel, did see a steep decline at the March 2022 breakpoint, after the rush of purchases before the fees, in line with the expected effects of the CCD fees in reducing the popularity of these vehicles. The large negative level shift for petrol vehicles in January 2020 most likely reflects the COVID-19 lockdown. However, it is very interesting to observe that though the EV registrations saw a drop after December 2023, the petrol and diesel vehicle registrations did not seem to pick up as before. This suggests that, although, there was a short-term effect when penalties were introduced, there was no long-term effect on the market for these vehicles, which includes the popular Utes. They retained their popularity even with the surcharge. It might have also played a role that the additional costs due to the CCD were relatively small compared to the overall costs of a Ute in New Zealand.

The level shifts here account for the jumps and drops as an immediate effect of these structural breaks, though they do not entirely illustrate the impact on the rate of adoption pre- and post-structural breaks. These slope shifts help to evaluate not only the initial change but also the longer-term evolution after the policy changes. Hence, Table 9 provides a detailed overview of the vehicle adoption rates and their corresponding slope shifts, followed immediately by the identified structural changes in the monthly registrations.

| **Motor Technologies** | **Timeframe according to the Break dates** | **Registration Slopes (change in monthly registrations per year)** | **Break dates** | **Slope Shift** |
|---|---|---|---|---|
| Battery Electric Vehicles (BEVs) | January 2014 - July 2017<br>August 2017 - June 2021<br>July 2021 - July 2022<br>August 2022 - December 2023<br>January 2024 - December 2024 | 59.60<br>37.76<br>24.26<br>583.41<br>574.40 | July 2017<br>June 2021<br>July 2022<br>December 2023 | -21.83<br>-13.50<br>559.14<br>-9.00 |
| Plug-in Hybrid Vehicles (PHEVs) | January 2014 - October 2017 | 13.81 | October 2017 | 6.02 |
| | November 2017 - May 2021 | 19.83 | May 2021 | 254.92 |
| | June 2021 - March 2022 | 274.76 | March 2022 | -913.6 |
| | April 2022 - February 2023 | -638.83 | February 2023 | 1555.34 |
| | March 2023 - December 2023 | 916.50 | December 2023 | -820.08 |
| | January 2024 - December 2024 | 96.41 | | |
| Hybrid Vehicles | January 2014 - November 2018 | 149.01 | November 2018 | 320.50 |
| | December 2018 - June 2020 | 469.52 | June 2020 | 2.66 |
| | July 2020 - March 2022 | 472.19 | March 2022 | -666.81 |
| | April 2022 - February 2023 | -194.61 | February 2023 | 4010.46 |
| | March 2023 - December 2023 | 3815.85 | December 2023 | -6187.91 |
| | January 2024 - December 2024 | -2372.05 | | |
| Petrol Vehicles | January 2014 - January 2020 | 98.35 | January 2020 | 2535.15 |
| | February 2020 - March 2022 | 2633.50 | March 2022 | -3490.34 |
| | April 2022 - December 2024 | -856.84 | | |
| Diesel Vehicles | January 2014 - May 2016 | 325.81 | May 2016 | -297.99 |
| | June 2016 - November 2019 | 27.82 | November 2019 | 757.52 |
| | December 2019 - February 2021 | 785.35 | February 2021 | 2016.97 |
| | March 2021 - March 2022 | 2802.32 | March 2022 | -3075.56 |
| | April 2022 – December 2024 | -273.24 | | |

**Table 9: Registration Slopes and Slope Shifts by Break Dates (2014–2024)**

In Table 9, the registration slopes represent the change in monthly registrations per year for the various motor technologies, capturing the longitudinal effects of the structural breaks. The slope shift here complements the level shifts of Table 8 and represents the prolonged effect of structural breaks through the change in momentum of the vehicle uptake. For the BEVs, the positive registration slopes correspond to a slow but constantly growing BEV adoption, which peaked from August 2022 to December 2023. The rate of BEV adoption does not seem to fall too much even after the withdrawal of the subsidies with a small negative slope shift, potentially reflecting the long-term impact of the policies on the consumer perception of the EVs. Differing from the BEV adoption rates, PHEVs and Hybrids exhibit a constantly growing positive adoption rate but a greater negative slope shift after December 2023. The negative registration slopes and slope shift for both PHEVs and Hybrids from April 2022 to February 2023, soon after the CCD implementation, might be caused by the growing preference for BEVs. As apprehended, the declining registration slopes and negative slope shift for the traditional fuel engine vehicles after March 2022 are consistent with a growing preference of consumers for EVs after the policy shifts. Overall, these slope shifts and level shifts highlight not only the momentary but also the sustained impact of the policy-driven structural changes in the adoption of these vehicles. The evolving trends and shifts in preferences favoring EVs, along with the constant decline in the uptake of internal combustion engine vehicles despite their popularity, suggest that targeted policy interventions can influence consumer preferences.

### 6.4 Policy Influence: A Difference-in-Differences (DiD) Approach

The following results are organized in the time frames depicting the introduction and discontinuation of the CCD scheme, to evaluate the effectiveness of EV incentives, as well as the impact on these registrations after the abrogation of the EV subsidies using a DiD regression model.

#### Impact of the Introduction of CCD Rebates

The incentives for EVs (Treatment) for New Zealand start in July 2021, while Queensland did not have any rebates at the state or federal level during that time. For the main analysis on log-transformed registrations, we consider July 2019 to June 2021 as the pre-treatment period and July 2021 to February 2022 as the post-treatment period. Additionally, visually analyzing the change in slope and the average EV registrations pre-EV incentive rollout provides valuable preliminary insights for a DiD regression analysis.

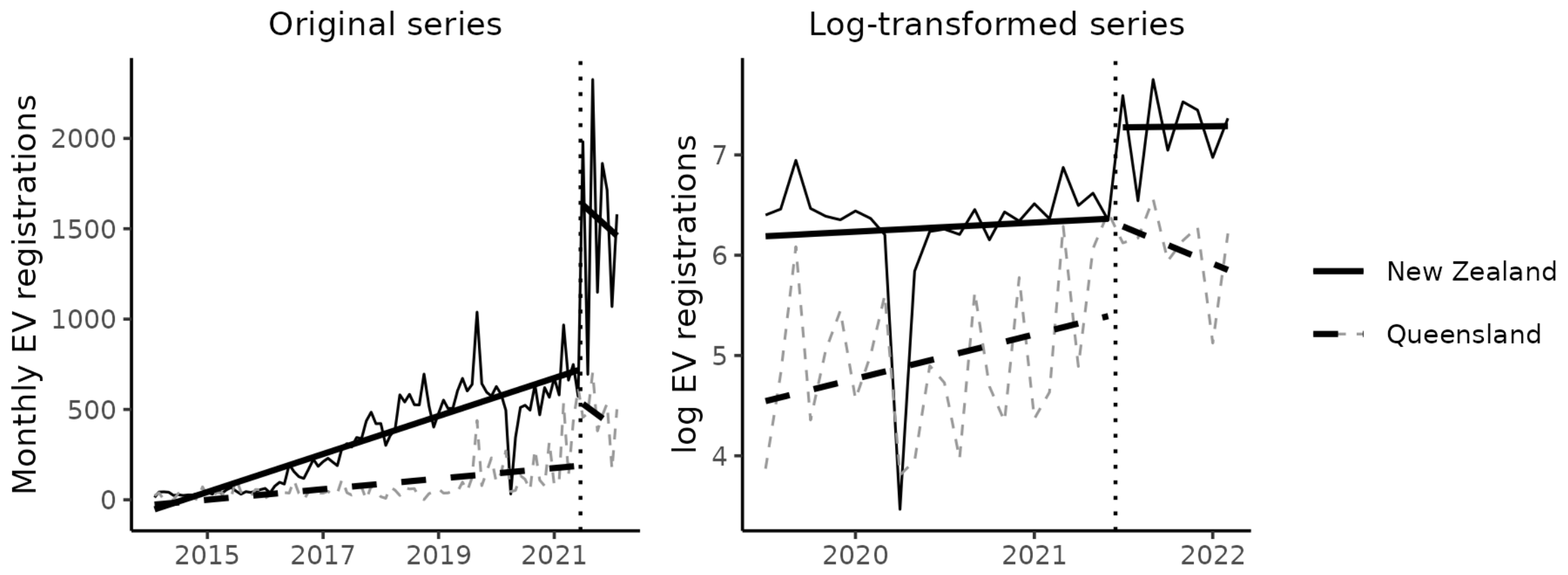

**Figure 6: Original and log-transformed time series of EV registrations for Queensland and New Zealand, with linear fits before and after July 2021 (original series: February 2014 to February 2022; log-transformed series: July 2019 to February 2022).**

| Region | July 2019 to June 2021 | July 2021 to February 2022 |
|---|---|---|
| Queensland (Control Group) | 4.97 | 6.07 |
| New Zealand (Treated Group) | 6.28 | 7.28 |

**Table 10: Average monthly log EV registrations in Queensland and New Zealand before and after the introduction of the CCD rebates**

In Figure 6, both New Zealand and Queensland show an upward trend in EV registrations before July 2021, with New Zealand having the steeper slope in the original series. After July 2021, EV registrations increased strongly in both regions. In the log-transformed series from July 2019, Queensland grew somewhat faster than New Zealand before July 2021, although the difference in slopes is not statistically significant ($p = 0.09$). After July 2021, the fitted line for Queensland declines, while the one for New Zealand is flat. Table 10 shows that the average log registrations increased by a similar amount in both regions. The DiD regression model quantifies this comparison.

The DiD regression on log-transformed registrations indicates that in the pre-treatment period the average log EV registrations were 4.97 in Queensland and 6.28 in New Zealand, corresponding to geometric means of about 144 and 532 registrations per month (Table 10). After July 2021, average log registrations increased by 1.10 in Queensland and by 1.00 in New Zealand, so that registrations roughly tripled in both regions. The DiD estimate is −0.10 (SE = 0.17; 95% CI [−0.43, 0.24]; $p = 0.58$), so there is no evidence that EV registrations in New Zealand grew by a larger proportion than in Queensland after the introduction of the rebates. Extending the post-treatment window to June 2022 gives a similar result (0.11; 95% CI [−0.30, 0.51]). Over the pre-treatment window, the log gap between the two regions shows a slight downward slope of −0.029 per month, which is not statistically significant at the 5% level ($p = 0.09$).

Allowing for a change in slope at the start of the policy, the log gap between New Zealand and Queensland widens by 0.093 per month after July 2021 in the main specification (95% CI [0.023, 0.163]), which uses the windows described in Section 5.4. To check the robustness of this result, the model was re-estimated with different windows (Table 11). Ending the pre-treatment window in December 2020 or March 2021 excludes the months immediately before the policy, when buyers in New Zealand may have postponed purchases until the rebates started. Starting the pre-treatment window in January 2018 uses a longer pre-treatment period. Ending the post-treatment window in December 2021 or January 2022 uses a shorter post-treatment period, and ending it in June 2022 includes the months in which purchases in Queensland were already eligible for its rebate. Across these variants, the estimated change in slope ranges from 0.07 to 0.13 per month. It is statistically significant at the 5% level in all variants except the one with the post-treatment window ending in December 2021, although only marginally when the pre-treatment window ends in March 2021. The two specifications rest on different assumptions about the counterfactual: without trends, the log gap would have remained constant, whereas with trends, Queensland would have continued to catch up. The results therefore range from no effect to a gradual acceleration of EV uptake in New Zealand relative to Queensland, and this evidence is suggestive rather than conclusive, given the short post-treatment window and the noisy Queensland series.

| **Specification** | **Change in slope per month [95% CI]** | **p-value** |
|---|---|---|
| Main: pre July 2019 to June 2021, post July 2021 to February 2022 | 0.093 [0.023, 0.163] | 0.009 |
| Pre-treatment window ends in March 2021 | 0.077 [−0.000, 0.154] | 0.050 |
| Pre-treatment window ends in December 2020 | 0.087 [0.001, 0.174] | 0.048 |
| Pre-treatment window starts in January 2018 | 0.122 [0.053, 0.192] | 0.001 |
| Post-treatment window ends in December 2021 | 0.070 [−0.026, 0.167] | 0.153 |
| Post-treatment window ends in January 2022 | 0.134 [0.045, 0.223] | 0.003 |
| Post-treatment window ends in June 2022 | 0.121 [0.078, 0.164] | <0.001 |

**Table 11: Change in the slope of the log gap between New Zealand and Queensland after July 2021 for different time windows. Newey-West standard errors.**

In untransformed registrations, the picture differs. Comparing February 2014 to June 2021 with July 2021 to February 2022, average monthly EV registrations rose from about 80 to 461 in Queensland and from about 332 to 1547 in New Zealand, and the naive DiD estimate is about 834 additional registrations per month. However, the gap between the two regions was already growing by about 6.3 registrations per month before July 2021 ($p < 0.001$), so the parallel trends assumption does not hold on this scale; extrapolating this pre-treatment trend reduces the estimate to about 527 registrations per month (95% CI [355, 700]). Because the New Zealand market was about four times larger than the Queensland market, equal growth rates are a more plausible counterfactual than equal absolute increases, which is why the log specification is preferred. Taken together, the results indicate that New Zealand's EV registrations rose strongly in absolute terms after the introduction of the rebates, but not by a larger proportion than in Queensland.

### Impact of the Discontinuation of CCD Rebates

The incentives for EVs under the CCD scheme ended in December 2023 for New Zealand. Meanwhile, Queensland also launched its ZEV rebate scheme in March 2022, with an increase in rebates in July 2023. As described in Section 5.4, the period from March 2022 to July 2023 was excluded because of the policy changes in both regions during that time. Furthermore, the structural break of December 2023 identified in the EV registration coincides with the repeal of the CCD scheme, indicating some influence of the termination on the EV uptake trends. A DiD regression model was fitted to the period from August 2023 to December 2024 to estimate the impact of ending the rebates in New Zealand. Thus, visually analyzing the change in slope and the average EV registrations post-EV incentive discontinuation provides valuable insights for a DiD regression analysis.

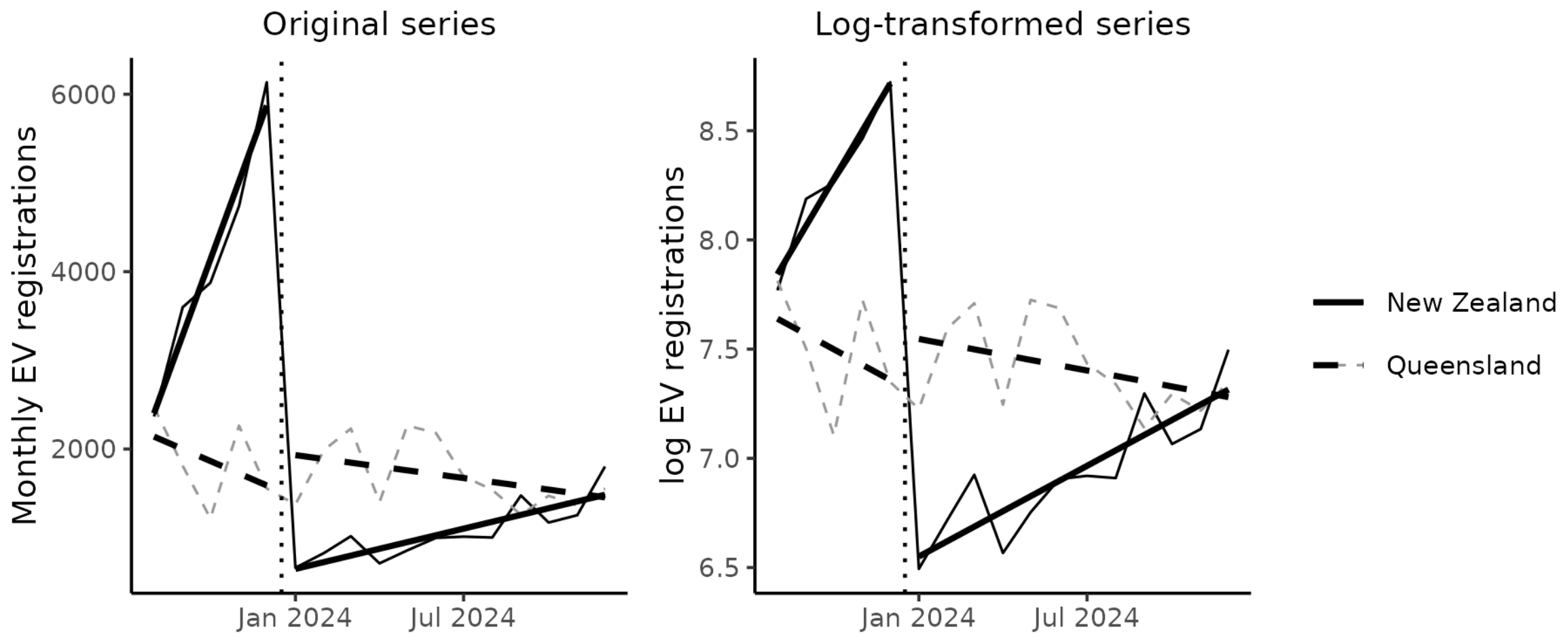

**Figure 7: Original and log transformed time series of EV registrations for Queensland and New Zealand with linear fits for August to December 2023 and January to December 2024.**

| Region | August to December 2023 | January to December 2024 |
|---|---|---|
| Queensland (Control Group) | 7.50 | 7.41 |
| New Zealand (Treated Group) | 8.28 | 6.93 |

**Table 12: Average monthly log EV registrations in Queensland and New Zealand in the pre-repeal and post-repeal windows**

In Figure 7 and Table 12, EV registrations in New Zealand increased strongly in the months before December 2023, as buyers brought purchases forward before the end of the rebates, and dropped sharply in January 2024. In contrast, the Queensland EV registrations were steadier, with no abrupt drop. During 2024, EV registrations in New Zealand recovered slowly, while those in Queensland declined slightly, but New Zealand remained well below its level of 2023 and below Queensland. The recovery may reflect an adjustment to the new pricing structure after the end of the rebates and the introduction of RUC, and possibly also an improved supply of EVs to New Zealand that developed during the CCD period. The DiD regression model quantifies these changes.

The DiD regression on log-transformed registrations indicates that in the pre-treatment period the average log EV registrations were 7.50 in Queensland and 8.28 in New Zealand (Table 12). After the termination of the rebates, average log registrations fell by 0.09 in Queensland but by 1.35 in New Zealand. The DiD estimate is −1.26 (SE = 0.28; 95% CI [−1.81, −0.72]; $p < 0.001$), corresponding to a decline of about 72% in New Zealand's EV registrations relative to Queensland. The pre-repeal window covers August to December 2023, including the final month of the rebates, and the post-repeal window covers January to December 2024. In untransformed registrations, the corresponding estimate is a decrease of about 2907 registrations per month (95% CI [−4423, −1390]).

A linear trend fitted to the log gap between New Zealand and Queensland over July 2021 to December 2023 yields a slope of −0.030 per month ($p = 0.03$). This test is of limited value, however. The window spans several policy changes in both regions (the start of Queensland's rebate in March 2022 and its increase in July 2023, the introduction of CCD fees in April 2022 and the CCD revision in July 2023), and the average slope reflects offsetting movements: the log gap widened by about 0.09 per month until mid-2022, narrowed by about 0.11 per month until July 2023, and widened sharply in the months before the repeal as buyers brought purchases forward. The repeal estimate should therefore be interpreted with caution, and part of it reflects the reversal of the pre-repeal rush rather than a persistent effect. The sharp decline in New Zealand's EV registrations after December 2023 nonetheless coincides with the withdrawal of the CCD incentives and the introduction of RUC, whose effects cannot be separated with these data. Registration data for 2025 suggest that the decline persisted: New Zealand recorded on average about 1,490 EV registrations per month, about a third more than in 2024 but less than half of the 2023 average.

### 6.5 EV Adoption: Impact on Carbon Footprints

The motivation for these policies was to reduce the transport-related emissions. Therefore, it is crucial to identify any significant deviations in the average CO2 emission levels around the policy evolution and shift in consumer preferences. Figure 8 depicts the monthly average exhaust CO2 emissions (g/km-3p WLTP) for initial registrations of vehicles every month, as published by the Ministry of Transport (2025b), along with any structural changes in these emissions from 2015 to 2024. Table 13 represents the structural breaks identified in this period.

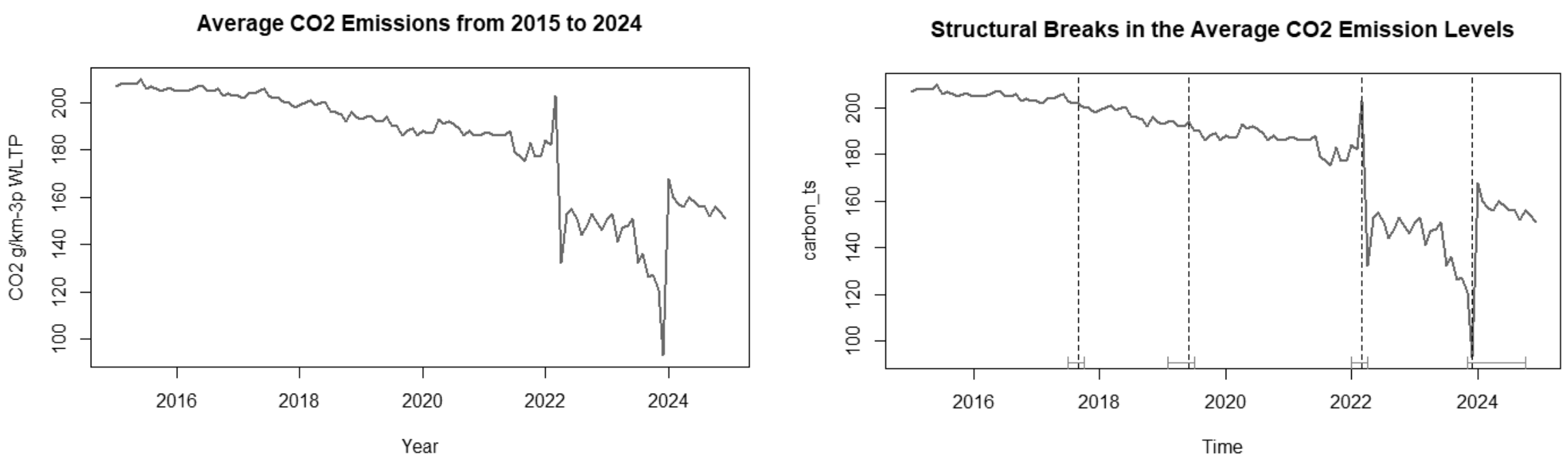

**Figure 8: Structural breaks in monthly average carbon emissions for initial registrations from 2015 to 2024.**

| Average Exhaust $CO_2$ emissions for Initial registrations (g/km-3p WLTP) | Break dates for Original Time Series (2015-2024) |
|---|---|
| Carbon Emissions | September 2017<br>June 2019<br>March 2022<br>December 2023 |

**Table 13: Break Dates in Average Exhaust CO2 Emissions for Initial Registrations (2015–2024)**

In Figure 8, two noteworthy structural breaks are identified that coincide with the policy timeline. In March 2022, average emissions jumped up, as buyers rushed to register high-emission vehicles before the fees started, and then dropped sharply in April 2022. In December 2023, immediately before the repeal of the CCD scheme, average emissions reached their lowest value, and they rebounded in January 2024. These breaks mark shifts in consumer decisions around the policy changes.

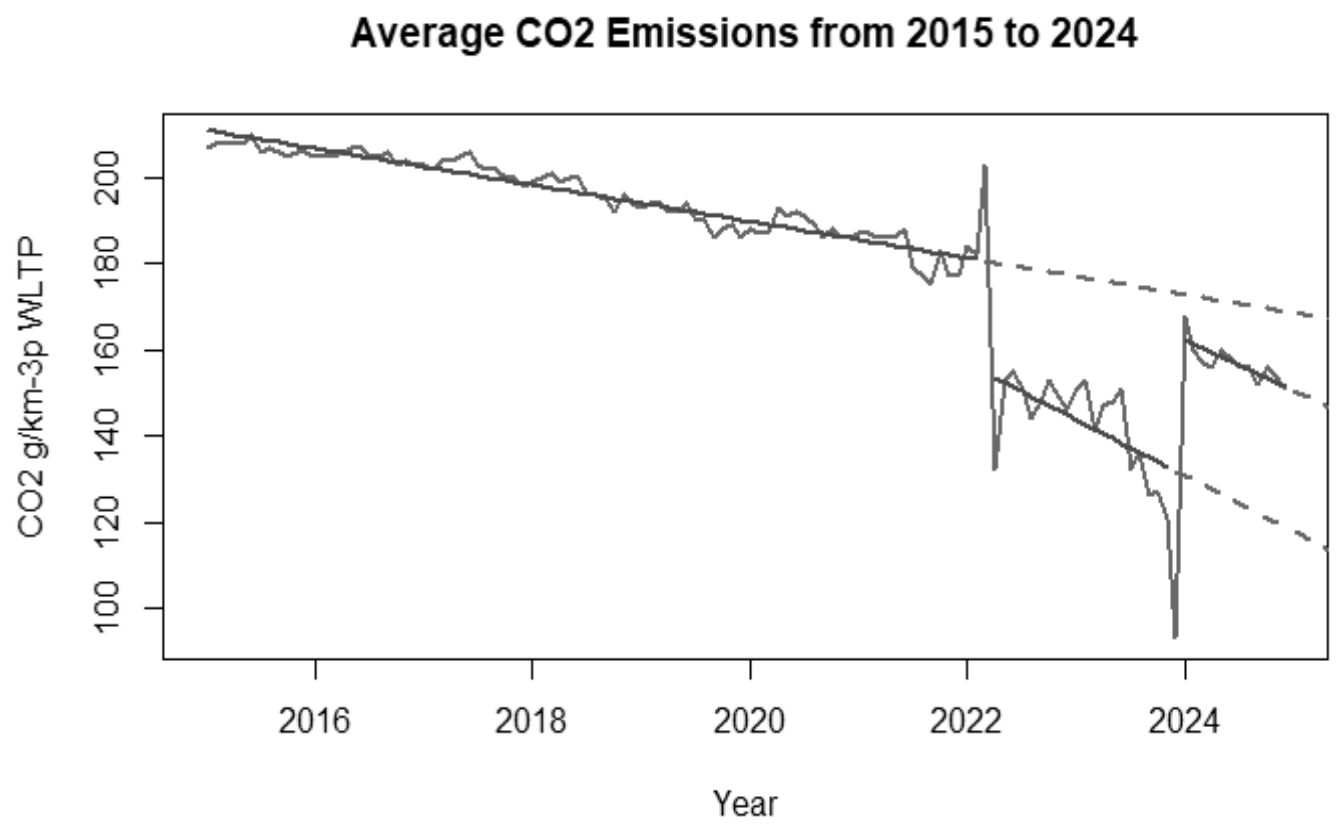

**Figure 9: Forecast for the monthly average carbon emissions for newly registered vehicles presents the predictable pattern post March 2022, December 2023, and December 2024 to observe the potential patterns each segment can project.**

| SEGMENTS | MONTH | ACTUAL AVERAGE CARBON EMISSIONS (g/km) | FORECASTED AVERAGE CARBON EMISSIONS (g/km) | ANNUAL CARBON EMISSIONS PER VEHICLE (tons) |
|---|---|---|---|---|
| January 2015-March 2022 | January 2015 | 207 | | 2.07 |
| | March 2022 | 203 | 180.54 | 1.81 |
| | December 2023 | | 173.12 | 1.73 |
| | December 2024 | | 168.88 | 1.69 |
| | December 2027 | | 156.15 | 1.56 |
| April 2022-December 2023 | April 2022 | 132 | | 1.32 |
| | December 2023 | 93 | 131.8 | 1.32 |
| | January 2024 | | 130.7 | 1.31 |
| | December 2024 | | 118.77 | 1.19 |
| | December 2027 | | 79.68 | 0.80 |
| January 2024-December 2024 | January 2024 | 168 | | 1.68 |
| | December 2024 | 151 | | 1.51 |
| | December 2027 | | 116.08 | 1.16 |

**Table 14: Actual and Forecasted Average Carbon Emissions and Annual Emissions per Vehicle by Segment (2015–2027)**

In Figure 9, the emissions from January 2015 to March 2022 exhibit a downward trend, though remaining at higher levels as compared to the actual emissions from April 2022 to December 2023. A notable reduction in carbon emissions is observed in December 2023, reaching the lowest values since 2015. The emissions rebounded in January 2024, yet stayed below the levels of March 2022. The forecast trends for segment 1 (January 2015-March 2022) further highlight that the downward trajectory continued for future years, although they were still not as low as the actual emissions for segment 2 (April 2022-December 2023). The forecasted values for segment 3 (January 2024-December 2024) are higher than the emission levels in December 2023, but still were considerably lower than those of the March 2022 forecasts. These trends are consistent with an influence of the CCD scheme on the average emissions of newly registered vehicles. The drops in the carbon emissions coincide with the peaks in the BEV and PHEV registrations, which suggests that the feebate scheme played a role in reducing carbon emissions. Facilitating a deeper insight into the evolution of emission patterns due to initial registrations, Table 14 depicts the forecasts for average annual emissions per vehicle registered in that month, when the average annual distance traveled per vehicle in New Zealand is 10,000 km, as reported by the Ministry of Transport (2015).

Table 14 reveals that the $CO_2$ emissions projected for Segment 2 saw a rapid decline, with the lowest predicted annual carbon emissions in December 2027. This underscores how a surge in EV fleet penetration can potentially reduce emissions. However, predictions for December 2027 for Segment 1 remained much higher, while the levels of annual emissions projected for Segment 3 were lower than those of Segment 1, yet were higher than those of Segment 2. This evolution of emission levels is consistent with an effect of the policy intervention. Without the policy changes, the emission levels might have followed the trend of Segment 1, with a much slower reduction. After December 2023, average emissions stayed below this trend. However, these extrapolations are mechanical and do not control for other factors, such as the global increase in EV supply, so they can only indicate that the policy interventions, even for a short time, contributed to the emission reduction in New Zealand.

## 7 Conclusion

This study examines how monetary benefits and financial penalties affected EV adoption in New Zealand between 2014 and 2024. The structural break analysis identified breaks that align with the timing of the policy interventions, in particular around the introduction and the end of the rebates for BEVs and PHEVs, and around the start of the fees for petrol and diesel vehicles, which was preceded by a rush of purchases in March 2022. The study also showed the evolution of the used and new BEV market in New Zealand, indicating changes in technological and market conditions over the past decade. When the fees ended in December 2023, there was no structural break in the registrations of petrol and diesel vehicles. Using the RDiT, we quantified the level shifts at the identified breaks. BEVs, PHEVs, and Hybrids experienced positive level shifts around the start of the CCD and its extension in 2022, and negative level shifts after the end of the CCD in December 2023, whereas petrol and diesel vehicles experienced negative level shifts in March 2022. In addition, a positive shift in the slopes was observed for BEVs in July 2022 and for PHEVs and Hybrids after February 2023, whereas Petrol and Diesel had a negative slope shift after March 2022. These breaks align with the policy changes, but by themselves they do not establish causal effects.

This study further examines, through a Difference-in-Differences approach with Queensland as a counterfactual, the effect of the EV incentives on EV adoption in New Zealand. The results are asymmetric. The introduction of the rebates coincided with a strong increase in EV registrations in New Zealand, but EV registrations in Queensland grew by a similar proportion, and the average effect on log registrations relative to Queensland is not statistically significant; there is suggestive evidence that the rebates gradually accelerated EV uptake. In untransformed registrations the increase is large, but the parallel trends assumption does not hold on that scale. The repeal of the rebates, in contrast, was followed by a decline of about 72% in New Zealand's EV registrations relative to Queensland, part of which reflects buyers bringing purchases forward before the rebates ended. For policy makers, the results suggest two lessons. First, the introduction of purchase rebates coincided with a period of strong EV growth in both regions, so that the additional effect of the rebates in relative terms is difficult to detect. Second, the abrupt end of the rebates, confirmed only a few weeks before it took effect, caused a rush of purchases followed by a sharp and persistent drop; a gradual and pre-announced phase-out might reduce such disruptions. Average $CO_2$ emissions of newly registered vehicles fell markedly during the CCD period and rebounded after its end, although part of the decline is likely also related to the global increase in EV supply.

Several limitations should be noted. First, the difference-in-differences design rests on a single control region. Obtaining monthly initial-registration data for jurisdictions with market conditions comparable to New Zealand proved challenging, and Queensland was ultimately the only feasible counterfactual; moreover, Queensland introduced its own ZEV rebate scheme for purchases from March 2022 and ended it in September 2024, within the post-repeal window, so the design lacks a cleanly untreated control throughout the full period. Second, with only two units, inference relies on Newey-West HAC standard errors rather than cluster-robust methods, and formal period-by-period event-study inference is not feasible. Third, the conclusions depend on the scale of the outcome. The parallel trends assumption is violated in untransformed registrations but plausible in log-transformed registrations, and the two scales lead to different conclusions about the introduction of the rebates. The log specification relies on a pre-treatment window of only two years and a post-treatment window of eight months, and the repeal estimate rests on a short pre-treatment window that is affected by buyers bringing purchases forward. Fourth, the two regions received treatments that were not identical in timing, eligibility, or strength, and New Zealand's contemporaneous penalties on high-emission vehicles, together with global supply and pricing dynamics over 2021 to 2023, may also have shaped the estimated effects. Finally, the carbon-emissions analysis relies on the average monthly exhaust intensity of newly registered vehicles rather than realized fleet-wide emissions, and the post-repeal window is short, limiting inference on long-run impacts.